\documentclass[12pt]{iopart}

\usepackage{graphicx}
\usepackage{caption}
\usepackage{subcaption}
\usepackage{float}
\usepackage{color}
\usepackage{amssymb}
\usepackage[numbers,square,sort&compress]{natbib}

\begin{document}
\title[Collective synchronization of dissipatively-coupled noise-activated processes]{Collective synchronization of dissipatively-coupled noise-activated processes}

\author{M Chatzittofi$^1$, R Golestanian$^{1,2}$  and J Agudo-Canalejo$^1$
}

\address{$^1$ Living Matter Physics, Max Planck Institute for Dynamics and Self-Organisation, D-37077, Goettingen, Germany}
\address{$^2$ Rudolf Peierls Centre for Theoretical Physics, University of Oxford, Oxford OX1 3PU, United Kingdom}
\ead{ramin.golestanian@ds.mpg.de; jaime.agudo@ds.mpg.de}

\begin{abstract}
A system of two enzymes mechanically coupled to each other in a viscous medium was recently studied, and conditions for obtaining synchronization and an enhanced average rate of the thermally-activated catalytic reactions of the enzymes were identified. The transition to synchronization occurred as the result of a global bifurcation in the underlying dynamical system. Here, we extend and generalize this idea to an arbitrary number of noise-activated cyclic processes, or oscillators, that are all coupled to each other \emph{via} a dissipative coupling. The $N$ coupled oscillators are described by $N$ phase coordinates driven in a tilted washboard potential. At low $N$ and strong coupling, we find synchronization as well as an enhancement in the average speed of the oscillators. In the large $N$ regime, we show that the collective dynamics can be described through a mean-field theory, which predicts a great enhancement in the average speed. In fact, beyond a critical value of the coupling strength, noise activation becomes irrelevant and the dynamics switch to an effectively deterministic ``running'' mode. Finally, we study the stochastic thermodynamics of the coupled oscillators, in particular their performance with regards to the thermodynamic uncertainty relation.
\end{abstract}

%\keywords{magnetic moment, solar neutrinos, astrophysics}
\submitto{\NJP}
\maketitle

\section{Introduction}

A system of driven, nonlinear coupled oscillators is nontrivial and can quickly lead to complex and unexpected situations when these oscillators synchronize, like in the famous case of the Millennium bridge \cite{strogatz2005crowd}. Generic features of synchronization have been widely studied using minimal models such as the Kuramoto model \cite{Kuramoto1984,pikovsky2007synchronization,acebron2005kuramoto,gupta2014kuramoto}. Networks of such oscillators give rise to fascinating phenomena such as states displaying coexistence of synchronization and incoherence, known as chimera states \cite{panaggio2015chimera}.

In biological systems, at the microscopic scale, the interactions are usually mediated by a viscous medium. For instance, hydrodynamic interactions can cause beating cilia or flagella to become synchronised  \cite{Golestanian2011,PhysRevLett.96.058102} displaying emergent phenomena such as metachronal waves \cite{Meng2021}. At an even smaller scale, on the scale of enzymes and molecular motors, many relevant processes are stochastic and thermally-activated: thermal noise is required to push the system over free energy barriers, e.g.~during chemical reactions inside enzymes or during the mechanical steps of molecular motors. These cyclic processes convert chemical energy into mechanical energy and heat \cite{PhysRevLett.72.2656, Golubeva_2012, PhysRevLett.109.168101}.

Enzymes and other catalytically active particles can self-organize in space thanks to the gradients generated by their nonequilibrium chemical activity \cite{ouazan2021non,cotton2022catalysis}. Additionally, the catalytic activity of the enzymes may be associated to conformational changes or oscillations in the enzyme shape \cite{glowacki2012taking, callender2015dynamical}. The effect of such conformational changes on the spatial dynamics and the rheology of enzyme-rich solutions has been a topic of great recent interest \cite{PhysRevLett.115.108102, hosaka2020mechanochemical, agudo2018enhanced, agudo2020diffusion}. In this context, a new mechanism for synchronization between two enzymes was recently reported, for enzymes that undergo conformational changes during their noise-activated catalytic steps, and which are coupled to each other through a viscous medium \cite{jaime}. The model for coupled phase dynamics that emerges after coarse graining the microscopic degrees of freedom in this system has some very peculiar features and emergent properties that are entirely different from those in conventional models for synchronization such as the Kuramoto model. In particular, interactions between phases are \emph{dissipative}, in the sense that they are mediated not by interaction potentials but rather by the mobility tensor (inverse to a friction tensor) that couples forces to velocities in the system. The same mobility tensor determines the stochastic noise in the system through the fluctuation-dissipation relation, making the model thermodynamically-consistent. Additionally, the transition to synchronization with increasing coupling was found to be due to a  global bifurcation in the underlying dynamical system, defined on the torus.

\begin{figure}
    \centering
    \includegraphics[scale=0.4]{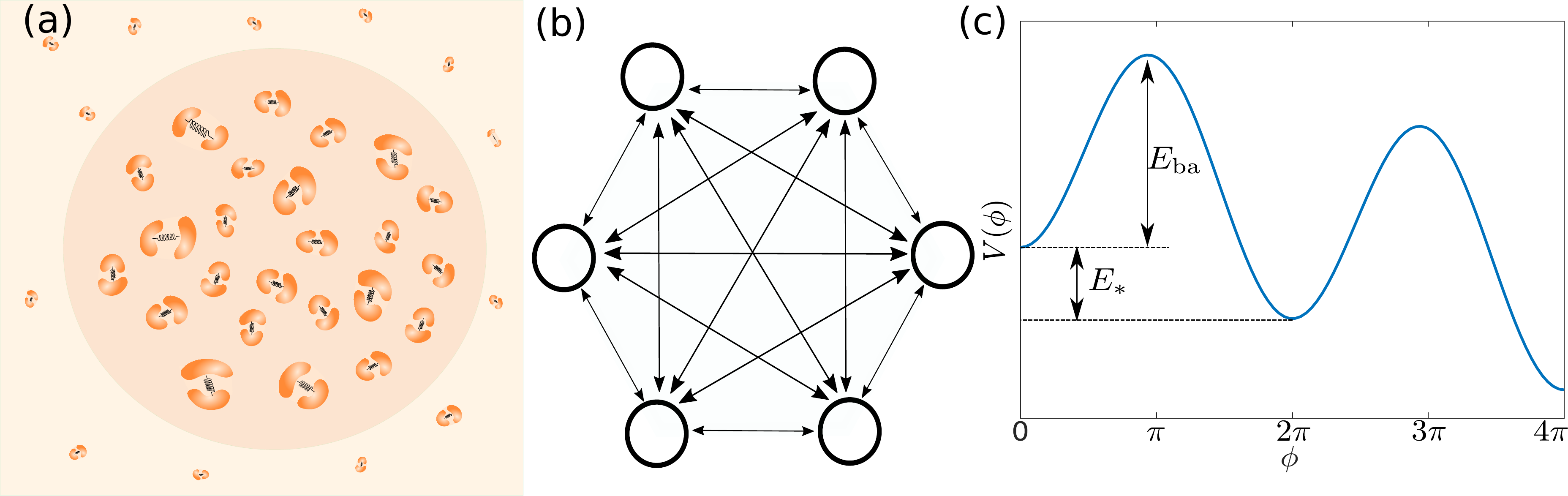} 
    \caption{(a) Enzymes densely clustered in a biomolecular condensate can mechanically interact with each other. (b) Network illustrating the ``all-to-all" interactions between the coupled oscillators considered here. (c) Each stochastic step corresponds to the phase $\phi$ advancing by $2\pi$ in a tilted washboard potential, which involves a noise-activated barrier crossing event.}
    \label{fig:intro}
\end{figure}

Inspired by these observations on the behaviour of two coupled enzymes, here we generalize the model to a system composed of an arbitrary number of stochastic oscillators, which interact with each other through a constant coupling of the dissipative kind. This could for example represent the interactions between enzymes in an enzyme-rich biomolecular condensate or metabolon \cite{sweetlove2018role,o2021role}, see figure~\ref{fig:intro}(a); but also any generic collection of noise-activated processes that are dissipatively coupled to each other. For simplicity, and in the spirit of minimal models of synchronization such as the original Kuramoto model \cite{Kuramoto1984}, we neglect the spatial structure and consider that each phase coupled with all other phases with equal strength, see figure~\ref{fig:intro}(b). The individual dynamics of each process is governed by a tilted washboard potential, see figure~\ref{fig:intro}(c). The resulting equations are rather generic and thus may find application as minimal models of not only catalytic processes but also other excitable systems \cite{lindner2004effects}, such as Josephson junctions \cite{wiesenfeld1989attractor,tsang1991dynamics,golomb1992clustering,watanabe1994constants} or firing neurons \cite{hodgkin1952quantitative,fitzhugh1961impulses,nagumo1962active}.

Because the model studied here is thermodynamically-consistent, it allows us to examine detailed features of its thermodynamic performance. A theoretical framework to understand the thermodynamics of fluctuating systems has been developed in recent years \cite{PhysRevLett.95.040602}. Of particular interest is a bound on the precision achievable by driven processes, determined by their entropy production, or equivalently their heat dissipation, known as the thermodynamic uncertainty relation \cite{PhysRevLett.114.158101,PhysRevLett.127.198101}. There is a growing interest in understanding how synchronization affects such thermodynamic measures of precision or efficiency, with applications to e.g.~beating cilia \cite{Hong2020}, generic Kuramoto oscillators \cite{Lee2018}, or molecular clocks \cite{Zhang2019}.

The paper is organized as follows. We begin by presenting the model of dissipatively coupled oscillators in its most general form, followed by its particularization to the minimal model studied here of $N$ identical oscillators with global (all-to-all) coupling. We then present the results of stochastic simulations for small and large numbers of oscillators. Next, we focus on the large $N$ limit, for which we show that the dynamics can be well understood using a mean-field theory. Finally, we study the stochastic thermodynamics of precision in the presence of coupling in our system.

\section{Model \label{sec:model}}

\subsection{Dissipative coupling}

We consider stochastic cyclic processes (oscillators) that are coupled to each other not through interaction forces, but through a mobility tensor that has nonzero off-diagonal components. That is, we will consider phases $\phi_\alpha$ with $\alpha=1,...,N$ which evolve according to the coupled overdamped Langevin equations
\begin{equation}
    \dot{\phi}_\alpha = - M_{\alpha \beta} \partial_\beta U + k_B T \Sigma_{\alpha \nu} \partial_\beta \Sigma_{\beta \nu} +  \sqrt{2 k_B T} \Sigma_{\alpha \beta}  \xi_\beta
    \label{eq:generalLangevin}
\end{equation}
where $\partial_\beta \equiv \frac{\partial}{\partial \phi_\beta}$, the Einstein summation convention for repeated indices is used, and the stochastic equation is to be interpreted in the Stratonovich sense. Here, the first term represents the deterministic velocity of the phases. The mobility tensor $M_{\alpha \beta}(\phi_1,...,\phi_N)$ can in principle be phase-dependent. For the dynamics to be thermodynamically consistent, this mobility tensor must be symmetric and positive definite \cite{degroot,kim2013microhydrodynamics}. Because there are no interaction forces between the phases, the potential $U$ is separable and may be written as $U(\phi_1,...,\phi_N)=V_1(\phi_1)+...+V_N(\phi_N)$. The third term represents the noise where, in order to satisfy the fluctuation-dissipation relation, $\Sigma$ is the square root of the mobility tensor defined \emph{via} $M_{\alpha\beta}=\Sigma_{\alpha \nu} \Sigma_{\beta\nu}$, and $\xi_\beta$ is unit white noise such that $\langle \xi_\beta(t) \rangle = 0$ and $\langle \xi_\alpha(t) \xi_\beta(t') \rangle = \delta_{\alpha \beta} \delta(t-t')$. The second term represents a spurious drift term that is only present when the noise is multiplicative, i.e.~when the mobility tensor is phase-dependent.

The stochastic dynamics given by (\ref{eq:generalLangevin}) may equivalently be written in the Fokker-Planck representation for the evolution of the probability distribution $P(\phi_1,...,\phi_N;t)$ as
\begin{equation}
\partial_t P = \partial_\alpha \left\{ M_{\alpha\beta} \left[ k_B T \partial_\beta P + (\partial_\beta U) P \right] \right\},
\label{eq:generalFP}
\end{equation}
which highlights that, when the choice of potential allows it, the system will relax to a steady state corresponding to thermodynamic equilibrium such that we recover the Boltzmann distribution $P(\phi_1,...,\phi_N) \propto \exp [ - U(\phi_1,...,\phi_N)/k_B T]$, independently of the choice of mobility tensor $M_{\alpha\beta}$. In fact, because the potential $U$ is separable, we may write $P(\phi_1,...,\phi_N)=\prod_{\alpha=1}^N P_\alpha(\phi_\alpha)$, with each phase independently satisfying the Boltzmann distribution $P_\alpha(\phi_\alpha)\propto \exp [ - V_\alpha(\phi_\alpha)/k_B T]$.

Importantly, when the choice of potential does not allow thermodynamic equilibrium, as in driven but periodic systems such as the ones that we will consider in the following, the system relaxes to a nonequilibrium steady state which (i) does depend on the choice of the mobility tensor $M_{\alpha\beta}$ and therefore on the strength of the coupling between oscillators determined by its off-diagonal components; and (ii) is no longer separable, so that there are correlations between the different phases.

A coupling of the form given by (\ref{eq:generalLangevin}) or equivalently (\ref{eq:generalFP}) arises naturally in processes that are coupled to each other through mechanical interactions at the nano- and microscale, as these are mediated by viscous, overdamped fields described by low Reynolds number hydrodynamics \cite{kim2013microhydrodynamics}. It represents a form of \emph{dissipative} coupling, as it can be understood as arising from taking the overdamped limit of full Langevin dynamics in the presence of a friction force on phase $\phi_\alpha$ going as $f_\alpha = - B_{\alpha \beta} \dot{\phi}_\beta$, where $B\equiv M^{-1}$ is a friction tensor.

\subsection{Noise-activated processes with global coupling}

As anticipated, we will consider here $N$ identical driven, noise-activated oscillators. This implies that the potentials for each phase are chosen to be identical, i.e.~$U(\phi_1,...,\phi_N)=V(\phi_1)+...+V(\phi_N)$, and $V(\phi)$ is chosen to be a tilted washboard potential of the form $V(\phi) = -F\phi - v\cos(\phi+\delta)$, with $F<v$ and $\delta = \arcsin(F/v)$ so that the minima are located at multiple integers of $2 \pi$. The values of $v$ and the driving force $F$ can be related to the energy barrier $E_{\mathrm{ba}}$ of the noise-activated step and to the energy $E_*$ released in each step, see figure \ref{fig:intro}(c), through $E_{\mathrm{ba}} = [2 \sqrt{1-(F/v)^2}-(F/v)(\pi - 2\delta)] v$ and $E_*= 2 \pi F$ \cite{jaime}. For an uncoupled oscillator, the height of the energy barrier relative to the thermal energy $k_B T$ controls the typical waiting time between stochastic steps, which scales as $e^{E_\mathrm{ba}/k_B T}$ for $E_\mathrm{ba} \gg k_B T$ \cite{kramers1940brownian}.  Note that, when $F>v$, the potential no longer displays energy barriers and becomes monotonically decreasing, so that the dynamics are no longer noise-activated.

We will consider the simplest possible dissipative coupling between the oscillators, where each of them interacts equally with all others \emph{via} a mobility matrix $M_{\alpha \beta}=\mu_\phi \tilde{M}_{\alpha \beta}$ with constant diagonal coefficients $\tilde{M}_{\alpha \alpha}=1$, and constant off-diagonal coefficients $\tilde{M}_{\alpha \beta}= h/(N-1)$ for $\alpha \neq \beta$. Here, $\mu_\phi$ sets the mobility scale, and $h$ is a dimensionless parameter that determines the strength of the coupling. This can be seen as an $N$-dimensional generalization of the two-dimensional problem studied in ref.~\cite{jaime}, with the additional simplification that the off-diagonal coefficients are constant, as it was shown in that work that this simplification does not affect the observed phenomenology. For the mobility matrix to be positive definite, the coupling strength must satisfy $-1 < h < N-1$. We will focus on positive values of $h$, for which the synchronization phenomenology is observed. Note that, since we choose the mobility matrix to be constant, the spurious drift term in the Langevin dynamics (\ref{eq:generalLangevin}) vanishes.

The deterministic ($k_B T=0$) version of the Langevin equation (\ref{eq:generalLangevin}) with the choice of washboard potential and mobility tensor just described can be written as $\dot{\phi}_\alpha = \mu_\phi \left\{ F - v \sin(\phi_\alpha+\delta) + \frac{h}{N-1} \sum_{\beta \neq \alpha} \left[ F - v \sin(\phi_\beta+\delta)  \right]  \right\}$. With the redefinitions $\omega\equiv (1+h)F\mu_\phi$, $a\equiv \mu_\phi v \left(1 - \frac{h}{N-1}\right)$, $K \equiv h \mu_\phi v \frac{N}{N-1}$, and $\theta_\alpha \equiv \phi_\alpha-\delta+\pi$, it may be rewritten as $\dot{\theta}_\alpha = \omega + a \sin\theta_\alpha + \frac{K}{N}\sum_{\beta} \sin \theta_\beta$. It is worth noting that this dynamical system has been previously studied in the context of superconducting Josephson junction arrays \cite{wiesenfeld1989attractor,tsang1991dynamics,golomb1992clustering,watanabe1994constants}. These studies however focused mostly in the barrier-free regime where no fixed points exist ($F>v$ or equivalently $\omega>a+K$), and mostly in its deterministic behavior, with a few exceptions in which an ad-hoc white noise was added \cite{wiesenfeld1989attractor,golomb1992clustering}. Here, on the other hand, we will focus on the stochastic dynamics of the system in the noise-activated regime ($F<v$ or equivalently $\omega<a+K$) where fixed points exist, and in the presence of a thermodynamically-consistent noise (satisfying the fluctuation-dissipation theorem). Furthermore we will focus on quantifying its collective, nonequilibrium stochastic dynamics (average speed, phase diffusion coefficient, phase correlations, and thermodynamic costs of precision), rather than on the properties of the underlying dynamical system.

\subsection{Quantitative measures of the stochastic dynamics}

In order to quantitatively assess synchronization, we must construct an order parameter. The usual order parameter in traditional synchronization problems, such as the Kuramoto model \cite{Kuramoto1984}, is $r\equiv\left\vert \frac{1}{N}\sum_{\alpha=1}^N e^{i\phi_\alpha} \right\vert$. However, in the context of noise-activated oscillators such as those studied here, the oscillators tend to spend a major fraction of the time in the stable fixed point corresponding to the minima of the driving potential (here at integer multiples of $2\pi$), independently of the strength of the coupling. The order parameter $r$ is thus not suitable in this context, as it would give $r \approx 1$ even for uncoupled oscillators.

Following ref.~\cite{jaime}, we will use the correlations between the stochastic dynamics of the oscillators as an order parameter. In particular, we may define the diffusion coefficient $D_\phi$ of an individual oscillator as
\begin{equation}
\langle (\phi_\alpha - \langle \phi_\alpha \rangle)^2 \rangle \sim 2 D_\phi t
\end{equation}
which is independent of $\alpha$, i.e.~equal for all oscillators. The $\sim$ symbol indicates that the equality is achieved asymptotically, at long times $t \to \infty$. We may also define the diffusion coefficient $D_\delta$ of the phase difference between a pair of oscillators, as
\begin{equation}
 \langle (\phi_\alpha - \phi_\beta)^2 \rangle \sim 2 D_\delta t
\end{equation}
which, due to the all-to-all coupling studied here, is identical for all pairs of oscillators $\alpha \neq \beta$. The two diffusion coefficients defined in this way are related to the correlation, or dimensionless covariance, between pairs of oscillators defined as
\begin{equation}
C = \frac{\langle (\phi_\alpha - \langle \phi_\alpha \rangle)(\phi_\beta - \langle \phi_\beta \rangle) \rangle}{\sqrt{\langle (\phi_\alpha - \langle \phi_\alpha \rangle)^2 \rangle \langle (\phi_\beta - \langle \phi_\beta \rangle)^2 \rangle}} = 1 - \frac{D_\delta}{2D_\phi}
\end{equation}
which we will use as our order parameter for synchronization, and is also identical for all pairs of oscillators $\alpha \neq \beta$. In particular, for uncorrelated oscillators we have $D_\delta = 2D_\phi$ and thus $C=0$, whereas for perfectly correlated oscillators we have $D_\delta=0$ and thus $C=1$. Anticorrelations would correspond to $C<0$, with a lower bound $C=-1/(N-1)$ for maximally anticorrelated oscillators. Lastly, we will define the average speed $\Omega$ of the oscillators as 
\begin{equation}
    \langle \phi_\alpha \rangle \sim \Omega t .
\end{equation}
which again is independent of $\alpha$.

\section{Results}

\subsection{Stochastic simulations}

\subsubsection{Small number of oscillators}

\begin{figure}%[ht]
     \centering
     \includegraphics[scale=0.35]{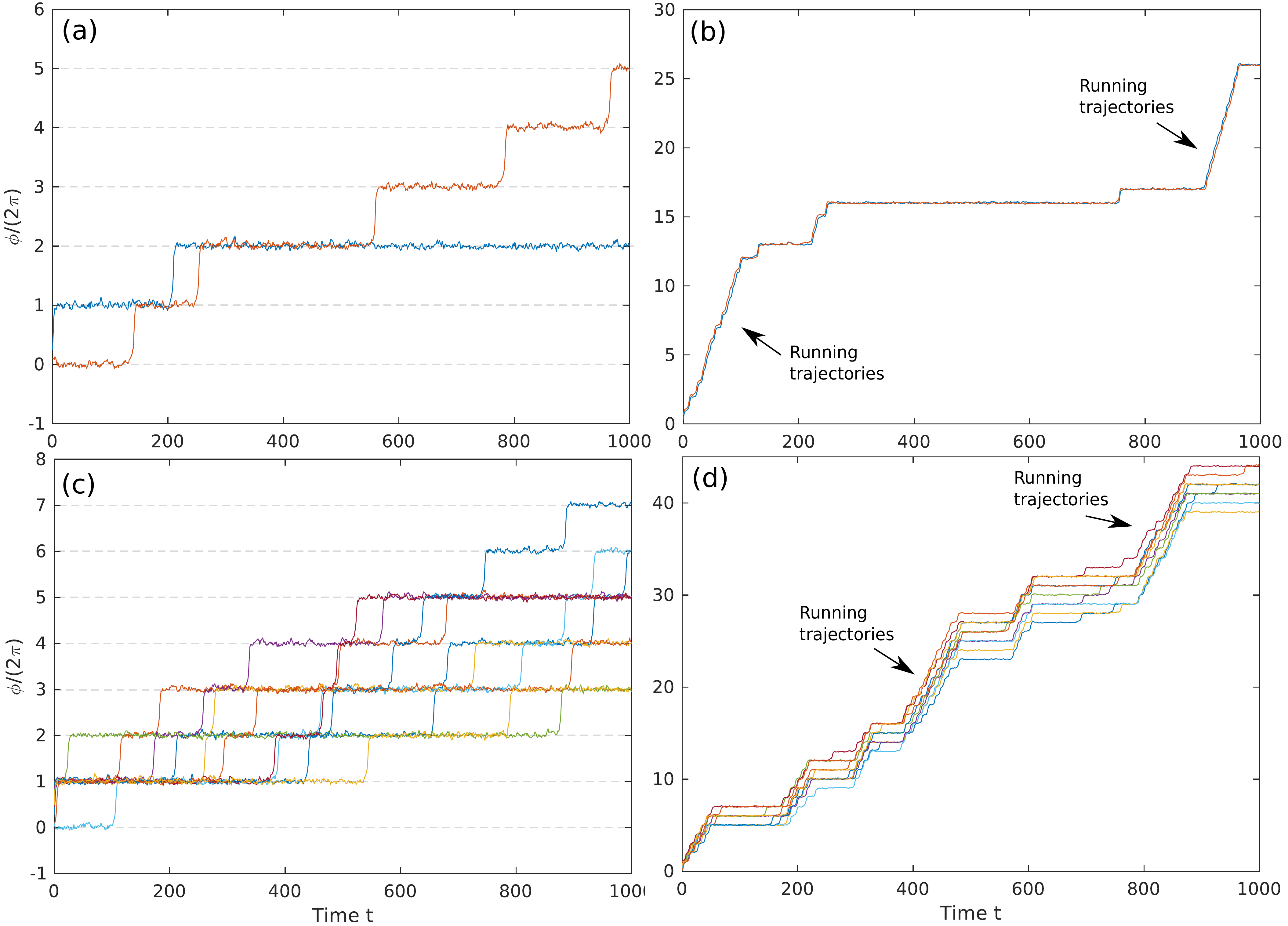}
     \caption{Stochastic trajectories for small number of oscillators, with $N=2$ in (a,b) and $N=10$ in (c,d). Without coupling [$h=0$ in (a) and (c)], the dynamics are clearly noise activated, with single steps occurring independently for each oscillator. With coupling [$h=0.5$ in (b) and $h=0.75$ in (d)], we observe strong correlations among all oscillators, and the dynamics moreover show multi-step ``runs'', marked by black arrows.  In all panels, $E_{\mathrm{ba}}/E_*=10^{-2}$ and $k_B T/E_{\mathrm{ba}} = 0.35$.}
     \label{fig:low_number_trajectories}
\end{figure}

For the case $N=2$ \cite{jaime}, we previously found that the system exhibits synchronisation and an enhanced average speed above a critical $h$. Examples of stochastic trajectories resulting from numerical solution of the Langevin dynamics (\ref{eq:generalLangevin}) for $N=2$ are shown in figure~\ref{fig:low_number_trajectories}(a)--(b). Note that, here and throughout the text, time is given in units of $(\mu_\phi v)^{-1}$. One clearly observes how, at zero or low coupling [figure~\ref{fig:low_number_trajectories}(a)], the trajectories appear independent, whereas at high coupling [figure~\ref{fig:low_number_trajectories}(b)] the two oscillators are strongly correlated. Moreover, the average speed increases: within the same timescale, a much larger number of steps is observed in the presence of coupling, and long ``runs'' of several consecutive steps are observed, as indicated by the black arrows. 

\begin{figure}%[ht]
    \centering
    \includegraphics[scale=0.58]{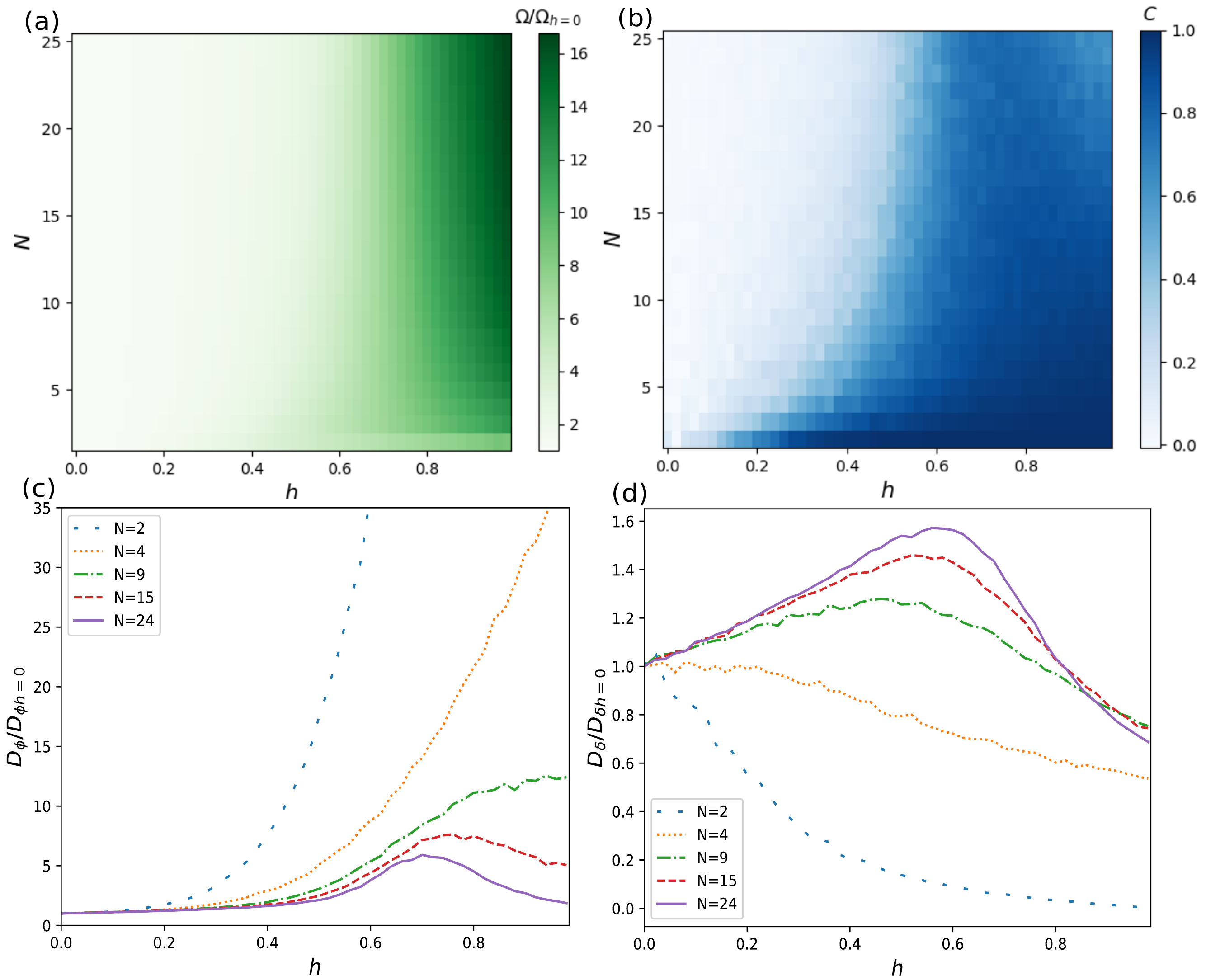} 
    \caption{Quantification of the dynamics for small number of oscillators. (a,b) Heatmaps of the average speed (a) and the synchronization order parameter (b), as a function of coupling strength and number of oscillators.  (c,d) Phase diffusion coefficient (c) and phase-difference diffusion coefficient (d) as a function of coupling strength, for several values of the enzyme number. Both are normalized by their value in the absence of coupling. In all panels, $E_{\mathrm{ba}}/E_*=10^{-2}$ and $k_B T/E_{\mathrm{ba}} = 0.5$}
    \label{fig:low_number_results}
\end{figure}

A similar behavior is observed for a larger, but still small, number of oscillators ($2<N\lesssim 25$). As seen in figure~\ref{fig:low_number_trajectories}(c)--(d), with increasing coupling the oscillators become correlated, the average speed increases, and long runs become apparent (black arrows). The dynamics are quantified in figure~\ref{fig:low_number_results} as a function of number of oscillators $N$ and strength of the coupling $h$. The average speed increases with increasing coupling, see figure~\ref{fig:low_number_results}(a), with this increase becoming significantly more pronounced at higher $N$. On the other hand, synchronization as measured by $C$ appears strongest and is present at smaller values of the coupling for lower $N$, see see figure~\ref{fig:low_number_results}(b). Interestingly, the phase diffusion coefficient $D_\phi$ appears to peak at a specific, $N$-dependent value of the coupling, see figure~\ref{fig:low_number_results}(c). The phase-difference diffusion coefficient $D_\delta$ also appears to peak at intermediate $h$ [figure~\ref{fig:low_number_results}(d)], but only for sufficiently large values of $N$, and to a much smaller extent than $D_\phi$, so that the order parameter $C$ still decreases with increasing $h$.

\begin{figure}%[ht]
    \centering
    \includegraphics[scale=0.27]{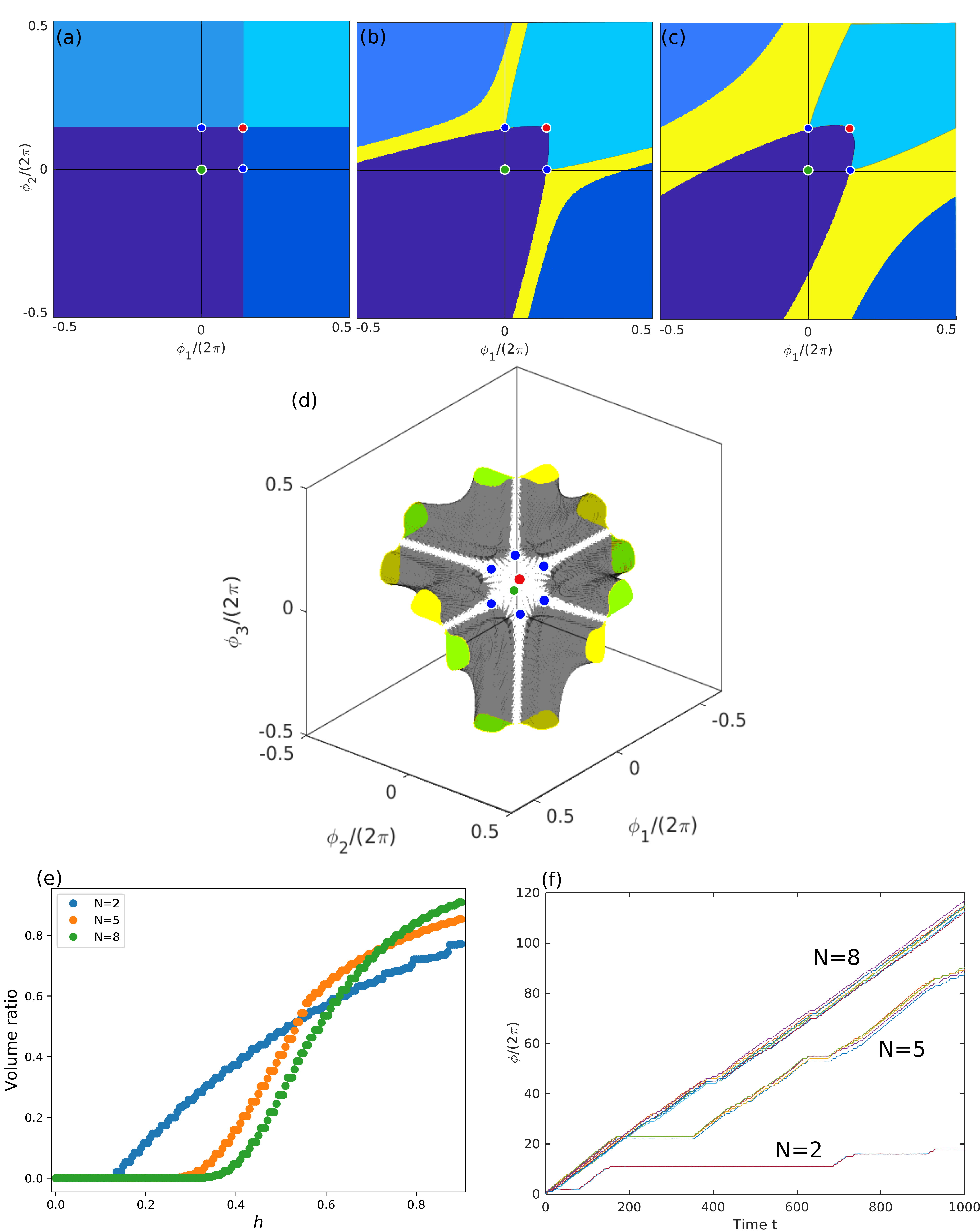} 
    \caption{(a)--(c) Phase portraits of the deterministic dynamics for $N=2$, with $E_{\mathrm{ba}}/E_*=10^{-2}$ and $h=0, 0.2, 0.5$ respectively. The differently-shaded blue regions correspond to basins of attraction of the stable fixed point at (0,0) which wind differently around the torus. The yellow region in (b,c) is the ``running band'' containing periodic orbits, which grows with increasing $h$ beyond the critical value \cite{jaime}. (d) The two ``running tubes'' containing periodic orbits that appear beyond the critical coupling for $N=3$, one corresponding to trajectories in which the phases advance in the (123) order (in green), the other in the (132) order (in yellow). Here, $E_{\mathrm{ba}}/E_*  =10^{-2}$ and $h=0.45$. In (a--d), the green, red, and blue circles represent the stable, unstable, and saddle fixed points of the dynamics. (e) The fraction of the phase space volume occupied by running bands as a function of the coupling strength, for several values of $N$. (f) Stochastic trajectories for the same values of $N$ when $h=0.6$ and $k_B T/E_\mathrm{ba} = 0.3 $. In (e,g), $E_{\mathrm{ba}}/E_*  = 2.5\times10^{-3}$. }
    \label{fig:volume_results}
\end{figure}

It is worth noting that enhanced phase diffusion in a tilted washboard potential has been previously reported for the motion of a single phase in such a potential \cite{PhysRevLett.87.010602,PhysRevE.65.031104}, in which case it was related to the transition from noise-activated dynamics to deterministic dynamics mediated by the saddle-node bifurcation of the system at $F=v$. However, in our case we have $F<v$ in all cases, i.e.~the dynamics remain noise-activated and integer values of $2\pi$ always correspond to stable fixed point of the $\phi_\alpha$'s, independently of the strength of the coupling $h$. A relation between these two distinct systems can still be established (and will be further clarified when we study the large $N$ limit in section~\ref{sec:meanfield}). In ref.~\cite{jaime}, exploring the case $N=2$, it was found that at a critical value of $h$ a global bifurcation of the underlying deterministic dynamical system ($k_B T=0$) occurs, giving rise to a splitting of the $(\phi_1,\phi_2)$ phase space, which corresponds modulo $2\pi$ to a torus, into two disconnected regions, see figure~\ref{fig:volume_results}(a)--(c). One region corresponds to the basin of attraction of the $(\phi_1,\phi_2)=(0,0)$ stable fixed point, whereas the new region (in yellow) corresponds to a band of periodic orbits, along which $\phi_1$ and $\phi_2$ increase deterministically. This bifurcation had been previously reported in a study of two coupled Josephson junctions \cite{tsang1991dynamics}. Our work further shows that this bifurcation is responsible for the emergence of synchronization, enhanced average speed, and the ``running trajectories'' in the stochastic system [$k_B T>0$, figure~\ref{fig:low_number_trajectories}(b) and (d)]. The enhancement in the phase and phase-difference diffusion coefficients observed at intermediate $h$ is therefore likely related to the transition from purely noise-activated dynamics before the bifurcation, to a mixture of noise-activated and deterministic dynamics once the periodic orbits have emerged. 

Indeed, an analysis of the deterministic dynamics for $N>2$ shows that a similar splitting of the phase space (now an $N$-torus) into disconnected regions occurs beyond an $N$-dependent critical value of $h$, one region corresponding to the basin of attraction of the fixed point, the others to periodic orbits along which all $\phi_\alpha$ increase deterministically. The regions containing periodic orbits in the case $N=3$ can be seen in figure~\ref{fig:volume_results}(d). In this case, there are two distinct ``tubes'' corresponding to periodic orbits in which $\phi_1$, $\phi_2$, and $\phi_3$ advance in the order (... 1 2 3 1 2 3 ...) and (... 1 3 2 1 3 2 ...), respectively. For arbitrary $N$, the number of higher-dimensional ``tubes'' containing periodic orbits therefore is $(N-1)!$, the number of circular permutations of $N$ distinct objects, as previously reported in the context of Josephson junction arrays \cite{wiesenfeld1989attractor}. We have measured the volume fraction of the phase space that is occupied by periodic bands as a function of $h$, for several values of $N$, see figure~\ref{fig:volume_results}(e). For this purpose, periodic trajectories were operationally defined as trajectories that, starting at $t=0$ from a point in the unit cell $-\pi<\phi_\alpha<\pi$ for all $\alpha$, reach $\phi_\alpha=6\pi$ for any $\alpha$ at some $t>0$, which implies that they do not end at the stable fixed point of the starting unit cell or any of its nearest neighbors. Beyond the critical $h$, the volume fraction grows from zero and saturates towards a limiting value as $h$ increases. Interestingly, with increasing $N$, the growth of this volume fraction beyond the critical $h$ becomes sharper, and the limiting value at large $h$ becomes closer to one. Extrapolating this trend we may speculate that, for large $N$, a sharp transition occurs at a critical $h$ at which the phase space becomes almost entirely occupied, or ``crowded'' \cite{wiesenfeld1989attractor}, by periodic orbits.
In figure~\ref{fig:volume_results}(f), some stochastic trajectories are shown, for $h$ beyond the critical value and several values of $N$. For $N=2$ we see longer periods in which the system is ``resting'' at the fixed point and the phases do not advance, interspersed with short deterministic runs. As $N$ is increased, the resting periods become shorter while the runs become longer, as one may expect from the considerations just described regarding the phase space volume occupied by periodic orbits.

\subsubsection{Large number of oscillators}

\begin{figure}%[ht]
    \centering
    \includegraphics[scale=0.34]{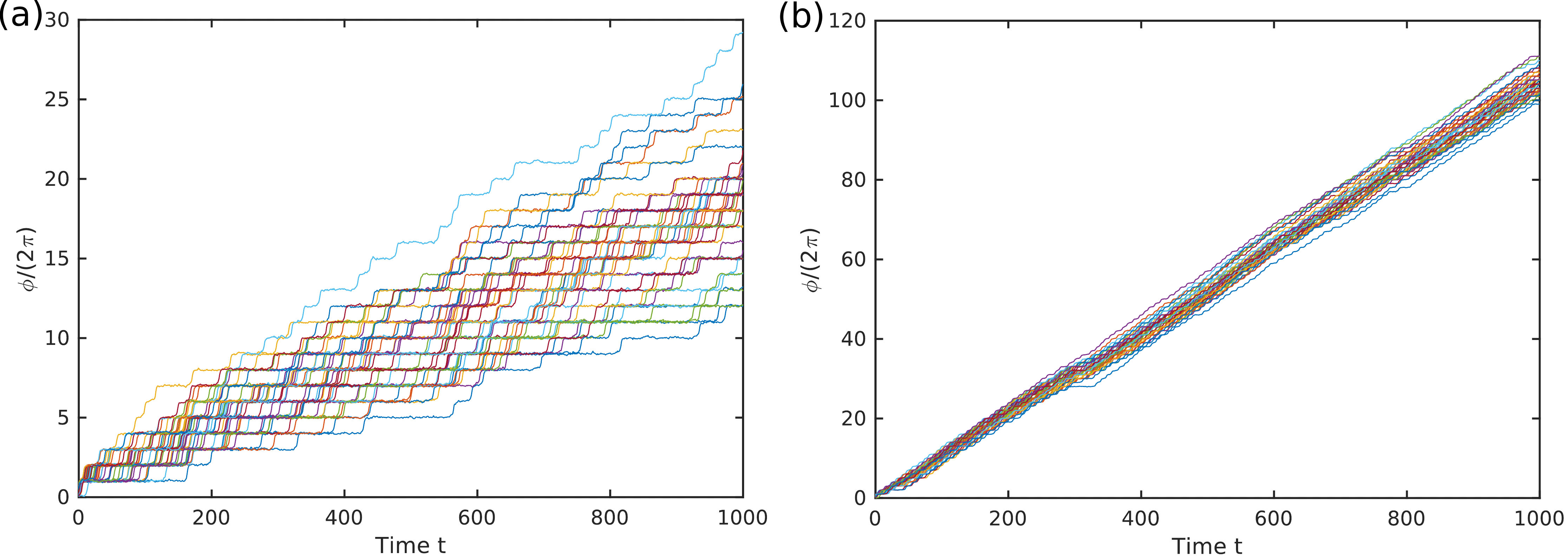} 
     \caption{Stochastic trajectories for small number of oscillators $N=50$. (a) For $h=0.5$, below the critical coupling, trajectories are largely uncorrelated, with single noise-activated steps. (b) For $h=0.8$, just above the critical coupling, trajectories are strongly correlated and mostly run deterministically. Here, $E_{\mathrm{ba}}/E_*=10^{-2}$ and $k_B T/E_{\mathrm{ba}}= 0.5$}
    \label{fig:many_enzymes_trajectories}
\end{figure}

\begin{figure}%[ht]
    \centering
    \includegraphics[scale=0.175]{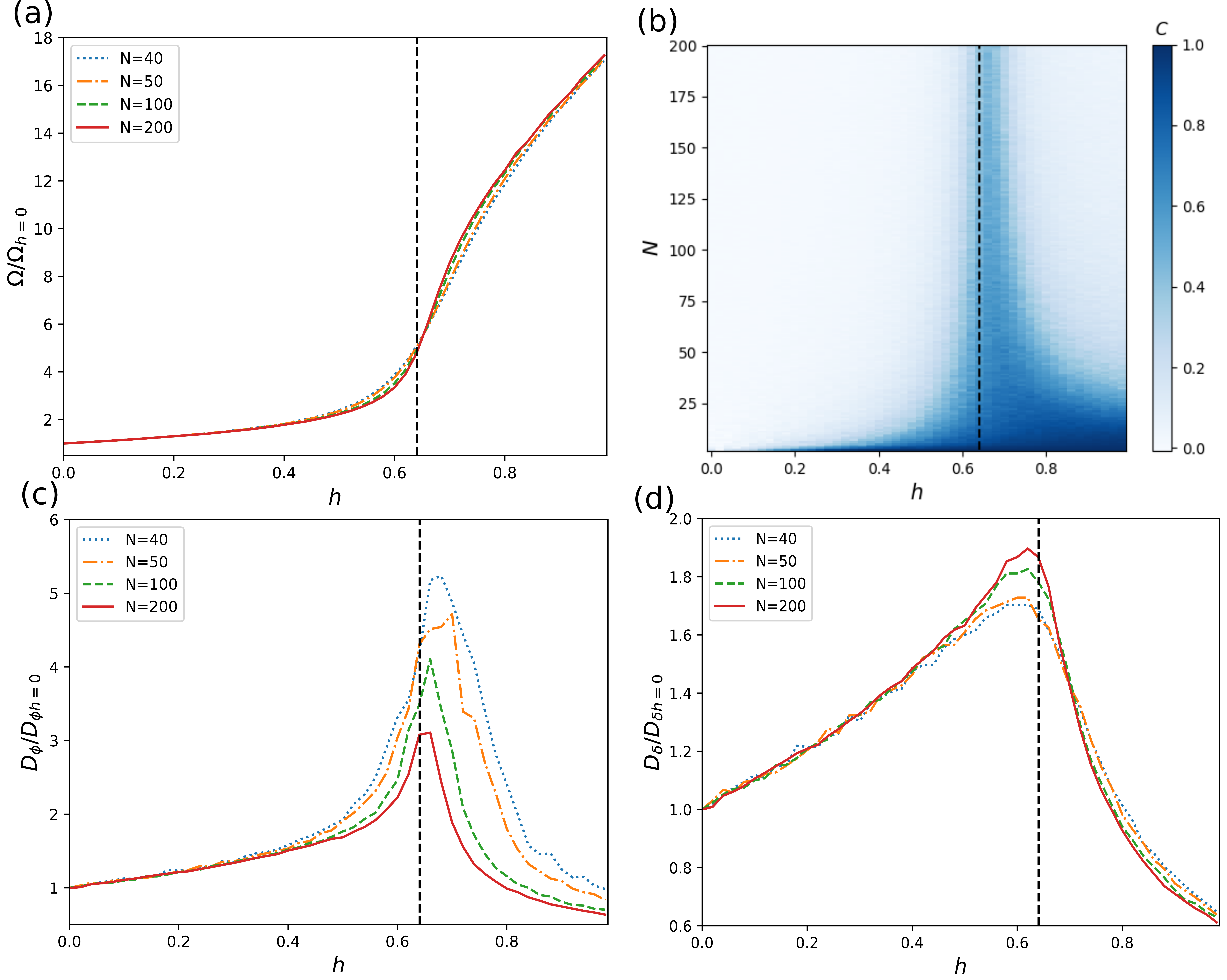} 
     \caption{Quantification of the dynamics for large number of oscillators. (a) Average speed as a function of coupling strength for several values of $N$. (b) Heatmap of the synchronization order parameter as a function of coupling strength and the number of oscillators. (c) Phase diffusion coefficient and (d) phase difference diffusion coefficient as a function of coupling strength for several values of $N$. Values in (a,c,d) are normalized by the value in the absence of coupling. In all panels, $E_{\mathrm{ba}}/E_*=10^{-2}$ and $k_B T/E_{\mathrm{ba}} = 0.5$, and the dashed black line corresponds to the critical coupling $h^* \approx 0.64$ predicted by the mean-field theory. }
    \label{fig:many_enzymes_results}
\end{figure}

As the number of oscillators increases, the behavior observed in stochastic simulations becomes largely independent of this number. We observe that, beyond an $N$-independent critical value of $h$, trajectories mostly run deterministically, without barely any resting periods at which the oscillators visit the fixed point of the dynamics, see figure~\ref{fig:many_enzymes_trajectories}. The stochastic dynamics are quantified for various values of $N$ in figure~\ref{fig:many_enzymes_results}. All relevant quantities $\Omega$, $D_\phi$, and $D_\delta$ depend only very weakly on $N$ and approach an asymptotic limit as $N \to \infty$, with $D_\phi$ showing the slowest approach towards this limit.

Interestingly, however, the synchronization order parameter becomes strongly nonmonotonic as a function of the coupling $h$, see figure~\ref{fig:many_enzymes_trajectories}(b): while synchronization is absent at low $h$, it rises sharply as we approach the critical $h$, but then quickly decreases back to zero (uncorrelated trajectories) as $h$ is further increased. Intuitively, in light of the results described in the previous section, this implies that the oscillators are most correlated the phase space volume fraction occupied by periodic orbits is intermediate, neither too small (in which case trajectories are predominantly noise-activated, with independent steps by each oscillator) not too large (in which case trajectories are effectively deterministic and uncoupled).

\subsection{Mean-field theory in the large $N$ limit \label{sec:meanfield}}

\subsubsection{General case}

The fact that the oscillators become uncorrelated in the large $N$ limit suggests that we may describe the behavior of the system through a mean-field theory. Then we begin from the Fokker-Planck equation of this model which, following (\ref{eq:generalFP}), is  given by
\begin{equation}
\partial_t P(\phi_1,...,\phi_N;t) = \partial_\alpha \left\{ \mu_\phi \tilde{M}_{\alpha\beta} \left[ k_B T \partial_\beta P + (\partial_\beta V(\phi_\beta)) P \right] \right\}.
\label{eq:particularFP}
\end{equation}
To study the large $N$ limit we first coarse grain over $(N-1)$ degrees of freedom to get an equation for the one-particle distribution $\rho(\phi;t)$, 
\begin{equation}
    \rho(\phi;t) = \int d \phi_2... d\phi_N P(\phi,...,\phi_N;t).
\end{equation}
By assuming that the processes are uncorrelated, so that the two-particle distribution reads $P_2(\phi,\phi_1;t) = \rho(\phi;t) \rho(\phi_1;t)$, we close the hierarchy of equations and obtain an equation for the one-particle distribution,
 \begin{equation}\label{eq:fplanck}
     \partial_t \rho(\phi;t) =\mu_\phi  \partial_\phi \left\{ k_B T \partial_\phi \rho- \left[F+hf_\mathrm{ave}-v \sin(\phi+\delta) \right]\rho \right\},
 \end{equation}
 where
  \begin{equation}\label{eq:fave}
     f_\mathrm{ave} = -\int^{2 \pi}_0 d \phi \rho(\phi;t) \frac{\partial V(\phi)}{\partial \phi} 
 \end{equation}
 is the average force experienced by an oscillator. Therefore, in the mean-field approximation each oscillator feels an effective driving force $F_{\mathrm{eff}}=F+hf_\mathrm{ave}$, independent of the number of enzymes, with the strength of the deviation from the true driving force $F$ governed by $h$. Notice that the equation of motion becomes nonlinear and nonlocal in $\phi$, due to the presence of $\rho$ in the definition of $f_\mathrm{ave}$.
 
The steady state $\rho_\mathrm{ss}(\phi)$ of the system can be found by imposing the condition of constant flux
  \begin{equation}
     - J/\mu_\phi = k_B T \partial_\phi \rho_\mathrm{ss} - \left[F+hf_\mathrm{ave}-v \sin(\phi+\delta) \right]\rho_\mathrm{ss},
     \label{eq:steadycond}
 \end{equation}
 where $J$ corresponds to the flux. This problem is identical to that of finding the steady state distribution of a single particle in a washboard potential with driving force $F_{\mathrm{eff}}=F+hf_\mathrm{ave}$, which is well studied and easily solved using standard methods \cite{risken1996fokker}, except that here one must additionally solve for $f_\mathrm{ave}$ in the implicit equation $f_\mathrm{ave} = \int^{2 \pi}_0 d \phi \rho_\mathrm{ss}(\phi) [F-v \sin(\phi+\delta)]$ (note that $\rho_\mathrm{ss}$ depends on $f_\mathrm{ave}$), required for self-consistency, see (\ref{eq:fave}). Armed with this self-consistent value of $f_\mathrm{ave}$ and thus of $F_\mathrm{eff}$, which is a function of all parameters of the system and in particular of the coupling $h$, we can then obtain the average speed $\Omega$ and the phase diffusion coefficient $D_\phi$ (which corresponds to $D_\delta/2$ given the absence of correlations) using the known results for a single particle in a tilted washboard potential \cite{risken1996fokker,PhysRevLett.87.010602,PhysRevE.65.031104}. Additionally, we may calculate a critical value of the coupling $h=h^*$ at which $F_\mathrm{eff}=v$, i.e.~the value of the coupling for which the energy barriers of the effective washboard potential vanish and the dynamics become deterministic (downhill). This further showcases the analogy between the giant diffusion observed at $F=v$ for a single particle in a washboard potential, and that seen at $h=h^*$ for both the phase and phase-difference diffusion coefficients in the present work.
 
\begin{figure}%[ht]
    \centering
    \includegraphics[scale=0.53]{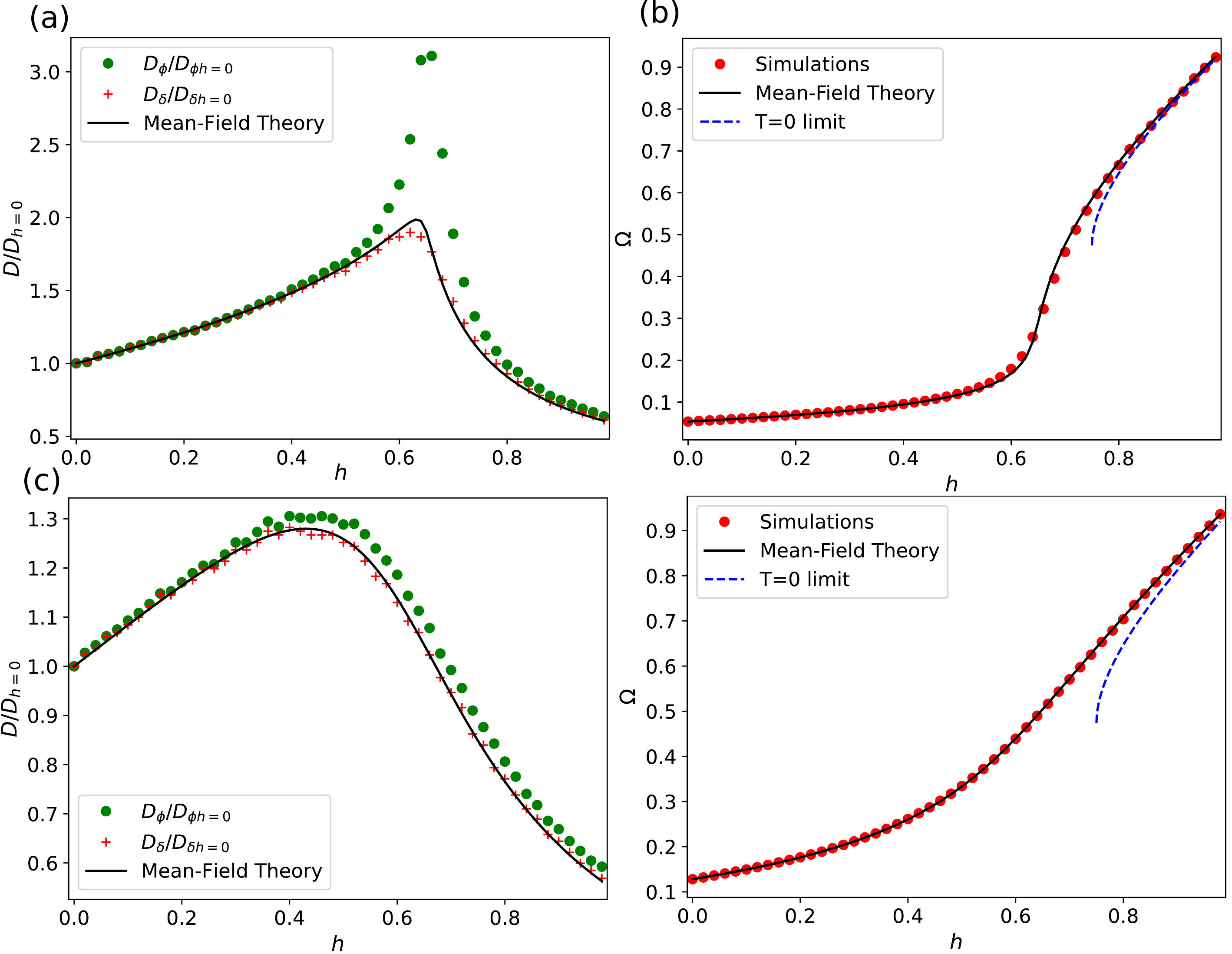} 
     \caption{Comparison between stochastic simulations for large number of oscillators ($N=200$) and the predictions of our mean-field theory. The strength of the noise is $k_B T/ E_{\mathrm{ba}} = 0.5$ in (a)--(b) and $k_B T/ E_{\mathrm{ba}} = 1$ in (c)--(d). In (a,c), the phase and phase-difference diffusion coefficients, normalized to their value in the absence of coupling, are shown as a function of coupling strength. In (b,d), the average speed (in units of $\mu_\phi v$) is shown as a function of coupling strength. The $T=0$ limit corresponds to equation~(\ref{eq:OmegaT0}). }
     \label{fig:mean_field_results}
\end{figure}

\begin{figure}%[ht]
    \centering
    \includegraphics[scale=0.7]{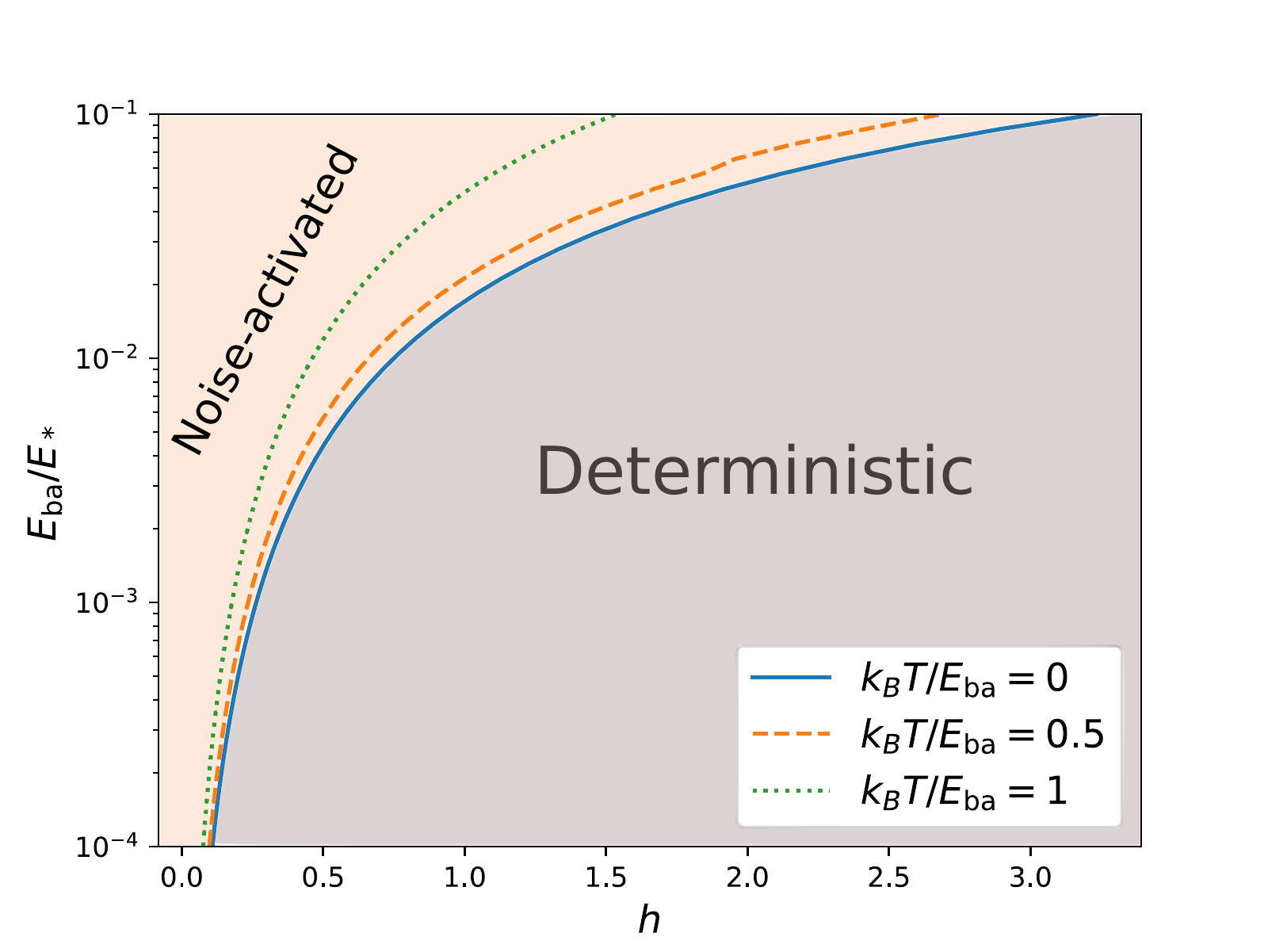} 
     \caption{``Phase diagram'' for the collective dynamics of a large number of dissipatively-coupled noise-activated processes, as a function of the coupling strength $h$ and the shape of the washboard potential $E_\mathrm{ba}/E_*$. The lines correspond to the critical coupling $h^*$ as calculated from the mean-field theory, for different values of the noise strength. The $T=0$ line (blue) corresponds to equation~(\ref{eq:crithT0}).}
     \label{fig:phasediag}
\end{figure}

 The values for $\Omega$, $D_\phi$, and $D_\delta$ obtained from this mean-field theory are compared to those obtained from stochastic simulations with $N=200$ in figure~\ref{fig:mean_field_results}, for two different values of the noise $k_B T$. We observe an excellent match, except for $D_\phi$ at the critical coupling which is underestimated (as described above, the limit $N\to \infty$ is approached slowly for $D_\phi$). Additionally, the critical coupling $h^*$ obtained from the mean-field theory is plotted as the vertical line in figure~\ref{fig:many_enzymes_results}, again with excellent agreement. As expected, it marks the transition between noise-activated and deterministic dynamics. In figure~\ref{fig:phasediag}, the critical coupling $h^*$ is shown as a function of the shape of the washboard potential [$E_\mathrm{ba}/E_*$, see figure~\ref{fig:intro}(c), which is in one-to-one correspondence with $F/v$], for various values of the noise strength. This defines a phase diagram marking the transition between noise-activated and deterministic dynamics.
 
 \subsubsection{Limit of vanishing noise \label{sec:meanfieldT0}}
 
Further analytical progress is possible in the limit of vanishing noise $T \rightarrow 0$. In this case, (\ref{eq:steadycond}) can be directly solved for $\rho_\mathrm{ss}$, giving
 \begin{equation}
     \rho_\mathrm{ss} (\phi) = \frac{1}{2\pi}\frac{\sqrt{(F+hf_\mathrm{ave})^2-v^2}}{F+hf_\mathrm{ave}-v\sin(\phi+\delta)}
     \label{eq:rhoss}
 \end{equation}
 where $J$ has been used to enforce normalization $\int_0^{2\pi}  \rho_\mathrm{ss} d\phi=1$. As expected, in the absence of noise the steady state is only well-defined when $F_\mathrm{eff}>v$ so that the potential admits deterministic dynamics. Using (\ref{eq:rhoss}) in the self-consistency condition gives $f_\mathrm{ave} = \sqrt{(F+hf_\mathrm{ave})^2-v^2}-hf_\mathrm{ave}$, and solving for $f_\mathrm{ave}$ and choosing the positive, physical root, we obtain
 \begin{equation}
     f_\mathrm{ave} = \frac{Fh + \sqrt{F^2(1+h)^2-(1+2h)v^2}}{1+2h}.
     \label{eq:fave2}
 \end{equation}
A real solution only exists when the term inside the square root is positive, which is possible when
 \begin{equation}
     h> \frac{v^2-F^2 + \sqrt{v^2-F^2}}{F^2} \equiv h^*
     \label{eq:crithT0}
 \end{equation}
 and serves to define the critical value of the coupling $h^*$ above which deterministic, ``running'' trajectories exist in the vanishing noise limit. Naturally, such a critical coupling is only well-defined when energy barriers are present in the true washboard potential ($F<v$), and $h^* \to 0$ from above as $F \to v$ from below. The critical coupling given by equation~(\ref{eq:crithT0}) is shown as the $T=0$ line in figure~\ref{fig:phasediag}, as a function of $E_\mathrm{ba}/E_*$ which (in one-to-one correspondence with $F/v$) defines the shape of the washboard potential.
 
 Interestingly, at the critical coupling we do not find $F_\mathrm{eff}(h^*)=v$ as one might have naively expected, but rather
  \begin{equation}
     F_\mathrm{eff}(h^*)=F+h^* f_\mathrm{ave}(h^*) = \frac{v^2}{F}
 \end{equation}
 which implies that $F_\mathrm{eff}(h^*)>v$ when $F<v$, i.e.~the effective washboard potential is already beyond the critical tilt, and the dynamics are therefore fully deterministic, when the critical coupling is reached. This is a reflection of the fact that the transition to the running state is discontinuous and noise-activated, as for $h>h^*$ the running state still coexists with the static state corresponding the fixed point $\phi_1=...=\phi_N=0$ (the transition state separating the two corresponds to the negative root of the self-consistency condition, with a minus sign in front of the square root of (\ref{eq:fave2})). This also implies that, at the critical coupling, there is already a finite, non-vanishing average speed $\Omega$ of the oscillators in the running state. Indeed, the average speed may be calculated as
 \begin{equation}
     \Omega = 2 \pi J =  \mu_\phi(1+h) f_\mathrm{ave}
     \label{eq:OmegaT0}
 \end{equation}
and, at the critical coupling, we find
 \begin{equation}
     \Omega(h^*) =  \mu_\phi(1+h^*) f_\mathrm{ave}(h^*) = \mu_\phi v \frac{  \sqrt{v^2-F^2}}{F}
 \end{equation}
 which is positive for any $F<v$. The zero temperature average speed as given by (\ref{eq:OmegaT0}) is plotted as the blue dashed lines in figure~\ref{fig:mean_field_results}(b,d). We see very good agreement with the stochastic simulations and the mean-field theory at finite temperature, serving as further confirmation that the dynamics beyond the critical coupling are largely deterministic.

\subsection{Stochastic thermodynamics of precision}

The coupling-induced transition marks a very strong change in the dynamics of the system, from noise-activated to deterministic. It is thus interesting to explore how does the transition affect the precision of the oscillators, which is bounded from below by the entropy production rate in the system \cite{PhysRevLett.95.040602}. Specifically, the thermodynamic uncertainty relation (TUR) states the bound \cite{PhysRevLett.114.158101}
 \begin{equation}\label{eq:TUR}
\dot{\sigma} t  \epsilon^2 \geq 2k_B ,
 \end{equation}
 where $\dot{\sigma}$ is the entropy production rate, and $\epsilon^2$ is the relative uncertainty defined as
  \begin{equation}\label{eq:relative_uncertainty}
     \epsilon^2 = \frac{\langle X^2 \rangle - \langle X \rangle^2}{\langle X \rangle^2},
 \end{equation}
 where $X$ is the observable of interest.
This inequality is crucial, since it implies that a higher precision in the catalytic rate (smaller $\epsilon$) requires higher entropy production or equivalently heat dissipation (higher $\dot{\sigma}$).
In our model the observable of interest is the total amount of phase advanced by the oscillators
\begin{equation}
    X = \sum_i^N \phi_i.
\end{equation}
The entropy production rate $\dot{\sigma}$ is easily calculated as
\begin{equation}\label{eq:entropy_production}
    \dot{\sigma} T = F \langle \dot{X} \rangle ,
\end{equation}
and is directly related to the free energy $E_* = 2 \pi F$ released in each noise-activated step, see figure~\ref{fig:intro}(c). 

 \begin{figure}%[H]
    \centering
    \includegraphics[scale=0.52]{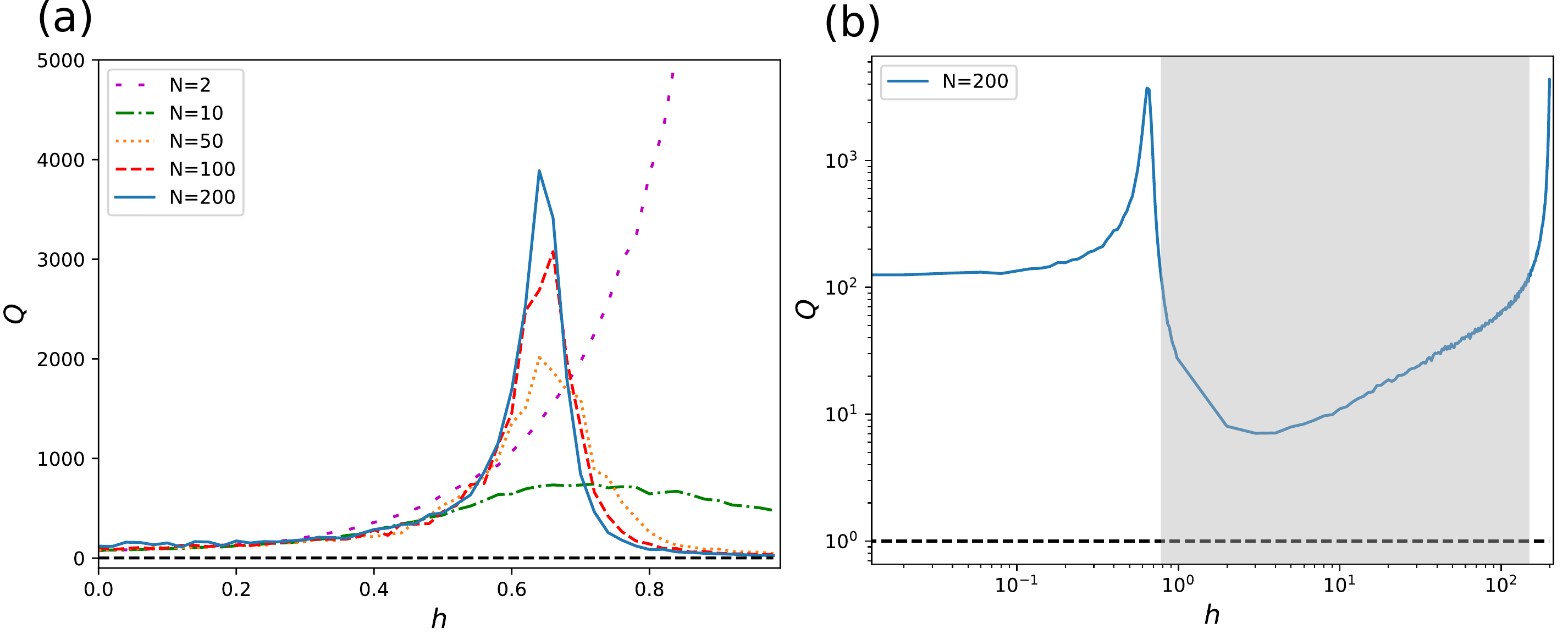} 
    \caption{Dimensionless thermodynamic uncertainty ratio $Q$ as measured in our stochastic simulations, as a function of coupling strength, for (a) weak coupling $0\leq h \leq 1$ and several values of the number of oscillators $N$; and (b) the full range of coupling $0 \leq h < N-1$ and $N=200$. The black dashed line corresponds to the lower bound $Q=1$ given by the TUR, which is satisfied for all $h$ as expected. For sufficiently large, but not too large values of the coupling, an enhancement in precision relative to that in the absence of coupling ($h=0$) is possible. This corresponds to the shaded region in (b). In both panels, $E_{\mathrm{ba}}/E_*=10^{-2}$ and $k_B T/E_{\mathrm{ba}}=0.5$.}
    \label{fig:tur_results}
\end{figure}

In figure~\ref{fig:tur_results}, we plot the dimensionless thermodynamic uncertainty ratio $Q \equiv \dot{\sigma} t  \epsilon^2 / (2k_B)$ as a function of the coupling $h$, as measured in our stochastic simulations for several values of $N$. According to the TUR (\ref{eq:TUR}), satisfies $Q \geq 1$. A process satisfying $Q=1$ is performing optimally from a thermodynamic perspective (as precisely as possible given its energy dissipation). We see that, as expected, the TUR is always respected. The behavior of $Q$ with increasing $h$ is strongly non-monotonic. Starting from the uncoupled case $h=0$, $Q$ first increases with increasing $h$, in the regime of noise-activated dynamics. After peaking around the critical $h=h^*$, however, $Q$ sharply decreases as $h$ is further increased and we venture further into the deterministic regime. Values of $Q$ smaller than the value $Q(h=0)$ are observed in this regime, implying that the coupling can enhance the thermodynamic performance of the oscillators. Finally, as we approach the upper bound of $h<N-1$ required by the positive definiteness of the mobility, $Q$ is observed to rise again. We note, however, that the regime $1 \ll h < N-1$ is somewhat artificial, as it corresponds to cases where the effects of cross-interactions between oscillators are much stronger than those of self-interactions. Moreover, this regime only exists for finite $N$ and becomes inaccessible in the thermodynamic limit $N\to \infty$.

\section{Conclusions}

We have studied a minimal model describing the collective dynamics of noise-activated cylic processes, or stochastic oscillators, that are coupled to each other through a dissipative coupling. That is, the processes are not coupled to each other through an interaction potential (or interaction force), but through the mobility tensor that connects forces to velocities in the overdamped dynamics. This mobility tensor also defines the properties of the stochastic noise through the fluctuation-dissipation relation, ensuring that the dynamics are thermodynamically-consistent and relax to thermodynamic equilibrium when such an equilibrium is available. Previously, we have shown how this kind of coupling arises naturally for processes that are mechanically coupled (e.g.~physically or through hydrodynamic interactions) in an overdamped, viscous medium \cite{jaime}.

For low $N$, where $N$ is the number of coupled oscillators, we found results analogous to those previously obtained for $N=2$ in ref.~\cite{jaime}. Beyond a critical coupling $h^*$, strong synchronization (as measured by the correlation function) and an enhancement in the average speed of the processes is observed. This transition can be understood as arising from a global bifurcation of the underlying dynamical system, defined on the $N$-torus, which leads to the emergence of periodic orbits that represent ``running'' trajectories, along which the phases of all the oscillators increase deterministically.

For large $N$, synchronization becomes confined to a narrow region near the critical coupling $h^*$. Below $h^*$, the dynamics are uncorrelated and stochastic (noise-activated). Above $h^*$, they are uncorrelated and effectively deterministic, and the average phase speed becomes greatly enhanced. We can understand this effect in two complementary ways: (i) analysis of the underlying dynamical system shows that, at large $N$, the volume fraction of the phase space occupied by periodic orbits increases very sharply at the bifurcation; and (ii) a mean-field theory shows that the energy barriers in the effective potential landscape experienced by each oscillator vanish at the critical coupling, and the dynamics become deterministic (downhill). The mean-field theory provides a great match to the results of stochastic simulations at large $N$ and allows for analytical prediction of the critical coupling and the average speed of the oscillators, particularly in the limit of low noise.

 Finally, we have shown that the oscillator dynamics can become more optimal in the presence of coupling, in the context of the trade-off between precision and entropy production described by the thermodynamic uncertainty relation. This occurs within the deterministic regime of the dynamics, beyond the critical coupling $h^*$. Over the full range of coupling strengths, the behavior relative to the thermodynamic bound on precision is rather complex, and signatures of the stochastic-to-deterministic transition at the critical coupling are clearly apparent in the precision.
 
Due to its simplicity, its general applicability to the description of coupled microscopic processes \cite{jaime}, and its intriguing features in the context of nonequilibrium statistical physics and dynamical systems theory, we believe that the model presented here merits significant further investigation.
Future work may consider local interactions between the nearest neighbours rather than all-to-all interactions as studied here, endowing the model with a spatial structure, as well as the role of quenched disorder \cite{Uchida2010}. Of great interest would also be the interactions between non-identical oscillators, as only the synchronization between identical oscillators has been studied so far. Lastly, one may explore connections to Bose-Einstein-like condensation in driven scalar active matter \cite{BECdriven} when interactions among particles are non-negligible.

\bibliography{bibv5}

\end{document}